\documentstyle[psfig,natbib209]{mn} 

\def\ifundefined#1{\expandafter\ifx\csname#1\endcsname\relax}

\def\la{\mathrel{\hbox{\rlap{\hbox{\lower4pt\hbox{$\sim$}}}\hbox{$<$}}}}
\def\ga{\mathrel{\hbox{\rlap{\hbox{\lower4pt\hbox{$\sim$}}}\hbox{$>$}}}}

\ifundefined{nuc}\def\nuc#1#2{\relax%
\ifmmode{}^{#1}{\protect\text{#2}}\else${}^{#1}$#2\fi}\else\relax\fi

\newcommand{\msol}{\ifmmode{{\rm M}_\odot}\else{M$_\odot$}\fi}
\newcommand{\foe}{\ifmmode{10^{51}}\else{$10^{51}$}\fi}

\newcommand{\xni}{\ifmmode{{\rm X}_{\rm Ni}}\else{X$_{\rm Ni}$}\fi}

\def\Teff{\ifmmode{T_{\rm eff}}\else{\hbox{$T_{\rm eff}$} }\fi}
\def\Rzero{\ifmmode{R_0}\else{\hbox{$R_0$} }\fi}
\newcommand{\vno}{\ifmmode{v_0}\else{\hbox{$v_0$} }\fi}

\newcommand{\be}{\begin{equation}}
\newcommand{\ee}{\end{equation}}

\newcommand{\eqs}[2]{Eqs. (\ref{#1}) \& (\ref{#2})}

\newcommand{\eq}[1]{Eq. (\ref{#1})} 
\newcommand{\fig}[2]{Figs  (\ref{#1}) \& (\ref{#2})} 
\newcommand{\Fig}[1]{Fig. (\ref{#1})} 
 \newcommand{\eqa}{\begin{eqnarray}}
\newcommand{\eeq}{\end{eqnarray}}  
\newcommand{\eqsto}[2]{Eqs. (\ref{#1}) to (\ref{#2})}

\newcommand{\heoh}{\ifmmode{{\rm [He/H]}}\else{[He/H]}\fi}
\newcommand{\heoo}{\ifmmode{{\rm He/O}}\else{He/O}\fi}
\newcommand{\coo}{\ifmmode{{\rm C/O}}\else{C/O}\fi}

\begin{document}

\bibliographystyle{natbib-apj}

\title{Inhomogeneous whistler  turbulence in  space plasmas}

\author[Dastgeer Shaikh]
  {Dastgeer ~Shaikh\thanks{email: dastgeer.shaikh@uah.edu} \\
   Department of Physics and Center for Space Plasma and Aeronomic Research (CSPAR)\\
The University of Alabama in Huntsville,
Huntsville. Alabama, 35899}

\maketitle

\begin{abstract}
A nonlinear two dimensional fluid model of whistler turbulence is
developed that nonlinearly couples wave magnetic field with electron
density perturbations. This coupling leads essentially to finite
compressibility effects in whistler turbulence model.  Interestingly
it is found from our simulations that despite strong compressibility
effects, the density fluctuations couple only weakly to the wave
magnetic field fluctuations.  In a characteristic regime where large
scale whistlers are predominant, the weakly coupled density
fluctuations do not modify inertial range energy cascade processes.
Consequently, the turbulent energy is dominated by the large scale
(compared to electron inertial length) eddies and it follows a
Kolmogorov-like $k^{-7/3}$ spectrum, where $k$ is a characteristic
wavenumber.  The weak coupling of the density fluctuations is
explained on the basis of a whistler wave parameter that quantifies
the contribution of density perturbations in the wave magnetic field.
\end{abstract}
\begin{keywords}
 (magnetohydrodynamics) MHD, (Sun:) solar wind, Sun: magnetic fields, ISM: magnetic fields
\end{keywords}

\section{Introduction}
Whistler wave regime is ubiquitously present in many space,
astrophysical and laboratory plasmas. For instance, in the
magnetospheric plasma, electrons can be accelerated by whistler-mode
and compressional ULF (fast mode waves) turbulences near the Earth's
synchronous orbit \cite[]{li2005}. The whistler-mode turbulence can
accelerate substorm injection electrons with several hundreds of keV
through wave-particle gyroresonant interaction and hence may play an
important role in the electron acceleration during substorms
\cite[]{li2005}. Vetoulis \& Drake (1999) describe whistler turbulence
at the magnetopause.  Whistler mode turbulence can be triggered by
electron beams in earth's bow shock \cite[]{tokar}. In the solar wind
plasma, observations have identified a spectral break in the solar
wind magnetic field spectrum (Goldstein et al 1995, Leamon et al
1999). The mechanism leading to the spectral break has been unclear
and thought to be either mediated by the kinetic Alfven waves
(Hasegawa 1976), or by electromagnetic ion-cyclotron-Alfven waves
\cite[]{yoon}, or whistler cascade regime \cite[]{gary}, or by a class
of fluctuations that can be dealt within the framework of the Hall
magnetohydrodynamic plasma model (Alexandrova et al 2007, 2008; Shaikh
\& Shukla 2008, 2009). Stawicki et al (2001) argue that Alfv\'en
fluctuations are suppressed by proton cyclotron damping at
intermediate wavenumbers so the observed power spectra are likely to
consist of weakly damped magnetosonic and/or whistler waves which are
dispersive unlike Alfv\'en waves. Moreover, turbulent fluctuations
corresponding to the high frequency and $k\rho_i \gg 1$ regime (where
$k$ is wavenumber, and $\rho_i$ is ion gyroradius) lead to a
decoupling of electron motion from that of ion such that the latter
becomes unmagnetized and can be treated as an immobile neutralizing
background fluid.  While whistler waves typically survive in the
higher frequency (and the corresponding smaller length scales) part of
the solar wind plasma spectrum, their role in influencing the inertial
range turbulent spectral cascades is still debated
\cite[]{biskamp,dastgeer03,dastgeer05,dastgeer08,dastgeer08a}.

Biskamp et al (1999) performed two and three dimensional simulations
of incompressible (density perturbations are ignored) electron MHD
model to demonstrate that the energy spectrum follows a $k^{-5/3}$ law
for $kd_e>1$ and $k^{-7/3}$ for $kd_e<1$. They further reported that
the 3D spectral properties are similar to those in 2D. This was lately
confirmed by Shaikh (2009) using 3D simulations. Cho \& Lazarian
(2004) performed 3D simulations of incompressible electron MHD model
to study anisotropic scaling that relates the parallel and
perpendicular wavenumbers through $k_\parallel \sim k^{1/3}_\perp$. In
a much detailed work, Cho \& Lazarian (2009) examined the anisotropy
in the electron MHD and showed that the high-order statistics in
electron MHD admit a scaling that is similar to the She-Leveque
scaling for incompressible hydrodynamic turbulence.

While there exists considerable literature describing the anisotropic
cascades in whistler turbulence, the role of whistler wave in the
spectral transfer of energy is still debated
\cite[]{biskamp,dastgeer03,dastgeer05,dastgeer08,dastgeer08a}.  For
instance, the Kolmogorov like dimensional arguments indicate that
propagation of whistlers in the presence of a mean or an external
constant magnetic field may change the spectral index of the inertial
range turbulent fluctuations from $k^{-7/3}$ to $k^{-2}$
\cite[]{biskamp}. By contrast, the numerical simulations
\cite[]{biskamp,dastgeer03,dastgeer05} suggest that whistler waves do
not influence the spectral migration of turbulent energy in the
inertial range despite strong wave activity and that the turbulent
spectra corresponding to the electron fluid fluctuations in whistler
turbulence continue to exhibit a Kolmogorov-like $k^{-7/3}$ spectrum.

What is not clear from these work is the quatitative role of whistler
and the corresponding mode coupling interactions that mediate the
inertial range turbulent spectra. Furthermore, much of the work
described above
\cite[]{biskamp,biskamp2,dastgeer03,dastgeer05,dastgeer08,dastgeer08a,cho1,cho2}
ignore the effect of density perturbations on the whistler mode
turbulence. It is unclear if density fluctuations in the electron
fluid modify the turbulent cascade properties.  Since density
fluctuations are critically important in many space and laboratory
plasma phenomena, their role in whistler turbulence needs to be
investigated.

The central object of this paper is to explore the nonlinear turbulent
processes mediated by whistler waves in the presence of density
perturbations. We will investigate nonlinear turbulent fluctuations,
based on nonlinear fluid simulations, in $\omega> \omega_{ci}$ regime
where correlation length scales of turbulence are comparable to the
electron inertial length scales.  Understanding of whistler turbulence
in the presence of density fluctuations is important in the context of
solar wind plasma
\cite[]{krafft,saito,Stawicki,gary,ng,Vocks,salem,Bhattacharjee1998},
magnetic reconnection in the Earth's magnetosphere \cite[]{Wei2007} to
interstellar medium \cite[]{burman} and astrophysical plasmas
\cite[]{roth} where characteristic fluctuations can typically be of
several astronomical units. These are only a few of the numerous other
studies. For more literature, the readers can refer to the simulation
work by \cite{biskamp} and others including Shukla (1978), Shukla et
al (2001), Galtier (2008), Urrutia et al (2008), Saito et al (2008),
Bengt \& Shukla (2008), Shaikh (2009), \cite{cho1,cho2} and numerous
references therein.

This paper is organized as follows. Section 2 describes nonlinear
whistler wave model for finite density perturbation. The finite
density perturbations are described in terms of plasma beta (ratio of
magnetic energy and electron fluid pressure). Section 3 deals with
nonlinear fluid simulations. Energy spectra are discussed in Section 4
and section 5 describes the effect of the density fluctuations in
whistler turblence. In section 6, we discuss spectral anisotropy in
inhomogeneous whistler turbulence.  Section 7 concludes our work.

\section{Whistler Wave  Model}
Fluctuations in a magnetized plasma excite whistler modes when they
propagate along a mean or background magnetic field with
characteristic frequency $\omega>\omega_{ci}$ and the length scales
are $c/\omega_{pi} < \ell < c/\omega_{pe}$, where $\omega_{pi},
\omega_{pe}$ are the plasma ion and electron frequencies.  In such a
high frequency regime, the ions do not have time to respond to the
electron motions.  Hence the electron dynamics plays a critical role
in determining the nonlinear interactions while the ions merely
provide a stationary neutralizing background against fast moving
electrons and behave as scattering centers. The whistler turbulence
can be described by the electron magnetohydrodynamics (EMHD) model of
plasma \cite[]{model} that deals with the single fluid description of
quasi neutral plasma.  The EMHD model has been discussed in
considerable detail in earlier work
\cite[]{model,biskamp,biskamp2,dastgeer00a,dastgeer00b,dastgeer03,dastgeer05,cho1,cho2}.
In whistler modes, the currents carried by the electron fluid are
important, and we therefore write down only those equations which are
pertinent to electron motion. These are electron fluid momentum,
electric field, currents, and electron continuity equations, \eqa
\label{elec}
m_e n_e \left(\frac{\partial}{\partial t} + {\bf V}_e \cdot \nabla  \right){\bf V}_e
=-en_e {\bf E} -  \frac{n_ee}{c} {\bf V}_e \times {\bf B} \nonumber \\
- \nabla \cdot {\bf P}
-\mu m_e n_e {\bf V}_e, 
\eeq
\be
{\bf E} = -\nabla \phi - \frac{1}{c} \frac{\partial {\bf A}}{\partial t},
\ee
\be
\label{ampere}
\nabla \times {\bf B} = \frac{4\pi}{c} {\bf J} +  
\frac{1}{c} \frac{\partial {\bf E}}{\partial t},
\ee
\be
\label{elec-den}
\frac{\partial n_e}{\partial t} + \nabla \cdot (n_e {\bf V}_e) = 0.
\ee Here $\nabla \cdot {\bf P} = \nabla P + \nabla \cdot {\bf \Pi}$,
the sum of pressure and stress tensors.  The pressure becomes highly
anisotropic in presence of a strong background magnetic field,
especially in the low beta solar wind plasma. Hence the total pressure
consists of the isotropic ($\nabla P$) and the anisotropic ($\nabla
\cdot {\bf \Pi}$) parts.  The remaining equations are ${\bf B} =
\nabla \times {\bf A}, {\bf J} = -en_e{\bf V}_e, \nabla \cdot {\bf B}
=0$. Here $m_e, n_e, {\bf V}_e$ are the electron mass, density and
fluid velocity respectively. ${\bf E}, {\bf B}$ respectively represent
electric and magnetic fields and $\phi , {\bf A}$ are electrostatic
and electromagnetic potentials. The remaining variables and constants
are, the collisional dissipation $\mu$, the current due to electrons
flow ${\bf J}$, and the velocity of light $c$.  The displacement
current in Ampere's law \eq{ampere} is ignored because
$\omega_{ce}/\omega_{pe} <1$.  The plasma is assumed to be
quasineutral, hence $n_e \approx n_i =n$. Density fluctuations are
considered to be incompressible at the leading order, but a first
order compressibility is included to describe finite electron plasma
beta (beta is the ratio of plasma pressure and magnetic field energy)
effects. We adopt the approach described by \cite{Abdalla} to include
the finite density perturbation effect in the whistler wave model that
couples density with pressure perturbations.  The density field is
determined from Poisson's equation and momentum equation as follows,
\be
\label{den-pert}
 -4\pi n_e e = \nabla \cdot {\bf E} = -\frac{1}{c} \nabla \cdot ({\bf V}_e \times {\bf B})
-\nabla \cdot \left( \frac{\nabla P}{n_e e} \right)
\ee
where electron inertia and pressure contribution is negligibly
small. Here perturbed density ($n_e$) is small compared to the mean
background density ($n_0$) such that $n_e/n_0<1$. Furthermore,
lengthscale on which density is varying ($\chi_n$) is small compared to
the characteristic length ($\ell$) i.e $\ell \chi_n <1$).
The leading order electron fluid velocity can
then be associated with the rotational magnetic field through
\be
\label{velocity}
{\bf V}_e = - \frac{c}{4\pi n_0 e}\nabla \times{\bf B}.
\ee
Substitution of  \eq{velocity} into \eq{den-pert} with further simplification gives
\be
\label{den-pert2}
n_e \approx \left(d_e\frac{\omega_{ce}}{\omega_{pe}} \right)^2 
\left( \hat{z} \cdot \nabla^2 \frac{\bf B}{B}_0 + \nabla^2\beta \right),
\ee
where $\hat{z}$ is the direction of the background magnetic field and
$\beta = 4\pi p/B_0^2$.  The above relation is further consistent with
$n_e/n_0<1$ since $\omega_{ce}/\omega_{pe} <1$ and ${\bf B}/{B}_0<1$
in the low beta whistler mode turbulence. This inequality may however
not hold for the $kd_e \gg 1$ modes.  In any event, the latter is
inaccessible by fluid theory as kinetic effects begin to play a crucial
role for the smallest scales. We nonetheless restrict ourselves to
$kd_e > 1$ only where the characteristic length scales are marginally
smaller than electron inertial length scales.  On taking the curl of
\eq{elec} and, after slight rearrangement of the terms, we obtain 
\eqa
\label{omega}
\frac{\partial {\bf \Omega}'}{\partial t} + 
\nabla \times ({\bf V}_e  \times {\bf \Omega}' ) = \nabla \times 
\left( \frac{\nabla \cdot {\bf P}}{m n_e} \right)
- \mu  \nabla \times{\bf V}_e,
\eeq
where
\[
{\bf \Omega}' = \nabla \times{\bf P} = d_e^2 \nabla^2 {\bf B} - {\bf B}.
\]
It can be seen from \eq{omega} that in the ideal whistler mode
turbulence (i.e. neglecting the term associated with the damping
$\mu$), the Curl of generalized electron momenta is frozen in the
electron fluid velocity.  This feature is strikingly similar to
Alfv\'enic turbulence where the magnetic field is frozen in the ideal
two fluid plasma \cite[]{biskamp03}.  Using electron continuity equation
\eq{elec-den} in combination with \eq{omega}, we obtain
\eqa
\left(\frac{\partial }{\partial t} + 
{\bf V}_e \cdot \nabla \right) \frac{{\bf \Omega}'}{n_e} = 
\left(\frac{{\bf \Omega}'}{n_e} \cdot \nabla \right)  {\bf V}_e +
\nabla \times 
\left( \frac{\nabla \cdot {\bf P}}{m n_e} \right) \nonumber \\
+\mu  \nabla \times{\bf V}_e,
\eeq
We next introduce normalized generalized vorticity ${\bf \Omega}$ as follows.
\[
{\bf \Omega} = \frac{{\bf \Omega}'}{n_e} \frac{n_0}{\omega_{ce}} = \frac{\bf B}{B_0}
-d_e^2 \nabla^2  \frac{\bf B}{B_0} -  \left(\frac{n_e}{n_0}-1 \right) \frac{\bf B}{B_0}
\]
On substituting ${\bf \Omega}'$ into \eq{omega} and using appropriate
vector identities, we obtain the three-dimensional normalized equation of EMHD
describing the evolution of the magnetic field fluctuations in
whistler wave,
\eqa
\label{emhd3}
\frac{\partial {\bf \Omega}}{\partial t} + {\bf
V}_e\cdot \nabla {\bf \Omega}-  {\bf \Omega}  \cdot \nabla{\bf V}_e 
 = 
\nabla \times 
\left( \frac{\nabla \cdot {\bf P}}{m n_e} \right) + \mu d_e^2 \nabla^2 {\bf B}.
\eeq 
The length scales in \eq{emhd3} are normalized by the electron skin
depth $d_e = c/\omega_{pe}$ i.e. the electron inertial length scale,
the magnetic field by a typical amplitude ${B}_0$, and time by the
corresponding electron gyro-frequency.  In \eq{emhd3}, the diffusion
operator on the right hand side is raised to $2n$. Here $n$ is an
integer and can take $n=1,2,3, \cdots$.  The case $n=1$ stands for
normal diffusion, while $n=2,3, \cdots$ corresponds to hyper- and
other higher order diffusion terms. \eq{emhd3} alongwith
\eq{den-pert2} form a complete set of three dimensional compressible
whistler wave model. For simulation purposes, we use two
dimensional (2D) model by ignoring the variation in the $z$=direction and
transforming the magnetic field as follows
\[
{\bf B}(x,y,t) = \hat{z}\times \nabla \psi(x,y,t) + \phi(x,y,t) \hat{z}.
\]
Such representation preserves $\nabla \cdot {\bf B} = 0$ in 2D.  Here
$\psi$ and $\phi$ are respectively orthogonal and logitudinal flux
functions of the perturbed magnetic field ${\bf B}$. This representation
transforms \eq{emhd3} into its parallel and perpendicular components (of the 
magnetic field) as follows  \cite{Abdalla},
\be
\label{2demhd-eq1}
\frac{\partial}{\partial t} (\phi-d_e^2\nabla \phi) = \frac{\partial n}{\partial t}
+[\phi, n+d_e^2\nabla^2 \phi ] - [\psi, \nabla^2 \psi ] + [\beta, n],
\ee
\be
\label{2demhd-eq2}
\frac{\partial}{\partial t} (\psi-d_e^2\nabla \psi) = 
-[\phi, \psi - d_e^2\nabla^2 \psi ] - [\beta, \nabla^2 \psi ],
\ee
\be
\label{2demhd-eq3}
\frac{\partial \beta}{\partial t} = -[\phi, \beta], 
\ee
\be
\label{2demhd-eq4}
n = \lambda^2 \nabla^2 (\phi+\beta),
\ee
where $\lambda = d_e{\omega_{ce}}/{\omega_{pe}}$.

The linearization of \eq{emhd3} about a constant magnetic field ${\bf
  B}=B_0\hat{z}+\tilde{\bf B}$, where $B_0$ and $\tilde{\bf B}$ are
respectively constant and wave magnetic fields, yields the following
equation,
\be
\omega_k (1+d_e^2k^2) \tilde{\bf B} + \frac{CB_0}{4\pi ne} ik_{\parallel} {\bf k} \times \tilde{\bf B}=0.
\ee
On eliminating the wave perturbed magnetic field from the above relation, one obtains
the following dispersion relation,
\be
\label{disp}
\omega_k = \omega_{ce}\frac{d_e^2 k_\parallel k}{1+d_e^2k^2},
\ee
where $ \omega_{ce}=eB_0/m_ec, k^2=k_x^2+k_y^2$ and $k_\parallel = {\bf k} \cdot {\bf B}_0$.
The use of \eq{disp} in  \eq{velocity} leads to the following relation
between the wave magnetic field and the velocity field,
\be
\label{bk}
\tilde{\bf B}= \pm \frac{i}{k}  {\bf k} \times \tilde{\bf B}
\ee

\begin{figure*}
\psfig{file=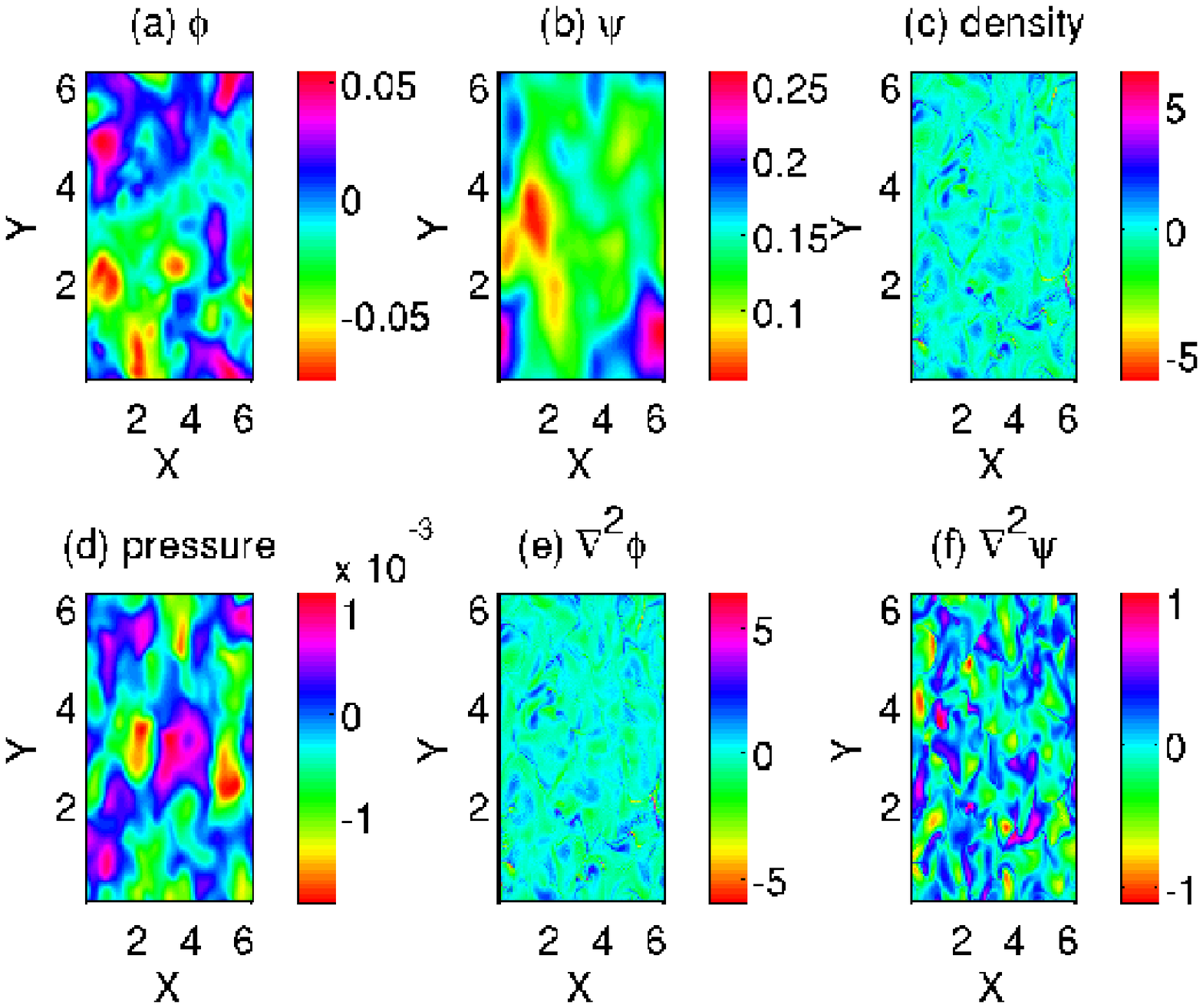,angle=0,width=0.9\textwidth}
\caption{\label{fig1}} 
Mode structures are shown in the low beta whistler turbulence
  simulations at a time when turbulence is fully saturated and is in
  the steady state. The effect of the background magnetic field is
  evident from these structures that are elongated along the $x$
  direction.  The simulation parameters are: Box size is $L_x \times
  L_y =2\pi \times 2\pi$, numerical resolution is $N_x \times N_y =
  512 \times 512$, electron skin depth is $d_e=0.015-1.0$, magnitude
  of constant magnetic field is $B_0=0.5$, dissipation $\mu=10^{-4}$,
  time step $dt=10^{-4}$.
\end{figure*}

The rhs of \eq{bk}, in combination with \eq{velocity}, corresponds
essentially to the whister wave perturbed velocity field. This
equation indicates that whistler waves consist of transverse
fluctuations in the magnetized space plasma and they are produced
essentially by rotational magnetic field that leads essentially to the
velocity field fluctuations. On replacing the rhs in \eq{bk} with the
perturbed velocity field, it can be shown that the whistler modes obey
equipartition between the magnetic and velocity field components as
$k^2|B|^2 \simeq |V_e|^2$. The whistler wave activity can thus be
quantified by how closely the characteristic modes obey the turbulent
equipartition relation.  In 2 and 3D cases, we have estimated this
relationship respectively in Shaikh \& Zank (2005) and Shaikh (2009)
in the incompressible limit.
Particularly interesting is the 2D case where \eq{bk} exhibits a
linear relation between $\psi_k$ and $\phi_k$ as $k|\psi_k|=|\phi_k|$
(Shaikh \& Zank 2005).  It is further evident from \eq{disp} that
there exists an intrinsic length scale $d_e$, the electron inertial
skin depth, which divides the entire turbulent spectrum into two
regions; namely short scale ($kd_e>1$) and long scale ($kd_e<1$)
regimes.  In the regime $kd_e<1$, the linear frequency of whistlers is
$\omega_k \sim k_y k$ and the waves are dispersive.  Conversely,
dispersion is weak in the other regime $kd_e>1$ since $\omega_k \sim
k_y/ k$ and hence the whistler wave packets interact more like the
eddies of hydrodynamical fluids.

\section{Nonlinear whistler turbulence}
We develop a two dimensional nonlinear fluid code to numerically
simulate \eqsto{2demhd-eq1}{2demhd-eq4} that describe low plasma beta
whistler turbulence.  The spatial descritization employs a
pseudospectral algorithm \cite[]{scheme,dastgeer06,dastgeer07} based on
a Fourier harmonic expansion of the bases for physical variables
(i.e. the magnetic field, velocity), whereas the temporal integration
uses a Runge Kutta (RK) 4th order method.  The boundary conditions are
periodic along the $x$ and $y$ directions in the local rectangular
region of the solar wind plasma.  The turbulent fluctuations are
initialized by using a uniform isotropic random spectral distribution
of Fourier modes concentrated in a smaller band of lower
wavenumbers. While spectral amplitudes of the fluctuations are random
for each Fourier coefficient, it follows a certain initial spectral
distribution proportional to $k^{-\alpha}$, where $\alpha$ is an
initial spectral index.  The spectral distribution set up in this
manner initializes random scale turbulent fluctuations.  We note that
a constant magnetic field is included along the $z$ direction
(i.e. ${\bf B}_0 =B_0 \hat {\bf x}$) to accommodate the large scale
(or the background solar wind) magnetic field.

The evolution of whistler fluid fluctuations are governed by the
nonlinear mode coupling interaction processes. In the presence of a
constant background magnetic field, turbulent fluctuations not only
couple nonlinearly with each other, but they also propagate along the
direction of the background magnetic field as small scale whistler
wave packets. The interaction of whistler waves with turbulent
fluctuations complicates the dynamical evolution. Additionally, by
virtue of nonlinear interactions the larger eddies transfer their
energy to smaller ones through a forward cascade. According to
\cite{kol}, the cascades of spectral energy occur purely amongst the
neighboring Fourier modes (i.e. local interaction) until the energy in
the smallest turbulent eddies is finally dissipated gradually due to
the finite dissipation. This leads to a damping of small scale
motions.  By contrast, the large-scales and the inertial range
turbulent fluctuations remain unaffected by direct dissipation of the
smaller scales. Since there is no mechanism that drives turbulence at
the larger scales in our model, the large-scale energy simply migrates
towards the smaller scales by virtue of nonlinear cascades in the
inertial range and is dissipated at the smallest turbulent
length-scales. A snap shot of fluctuations in density, magnetic and
pressure fields is shown in \Fig{fig1}. Consistent with 2D turbulence,
a dual cascade phenomenon is observed \cite{krai}.  In this process,
the perpendicular component ($\phi$) of the magnetic field cascades
predominantly towards the smaller scales whereas the parallel
component ($\psi$) exhibits large scales in its spectrum (see Fig 1a
\& 1b). By contrast, density and pressure fields comprise of smaller
scales (Fig 1c \& 1d) as they follow a forward cascade. Since the
spectrum of vorticity fields ($\nabla^2 \psi$ and $\nabla^2 \phi$) is
dominated by the large $k$ modes, they contain smaller scales (Fig 1e
\& 1f).

\begin{figure}
\psfig{file=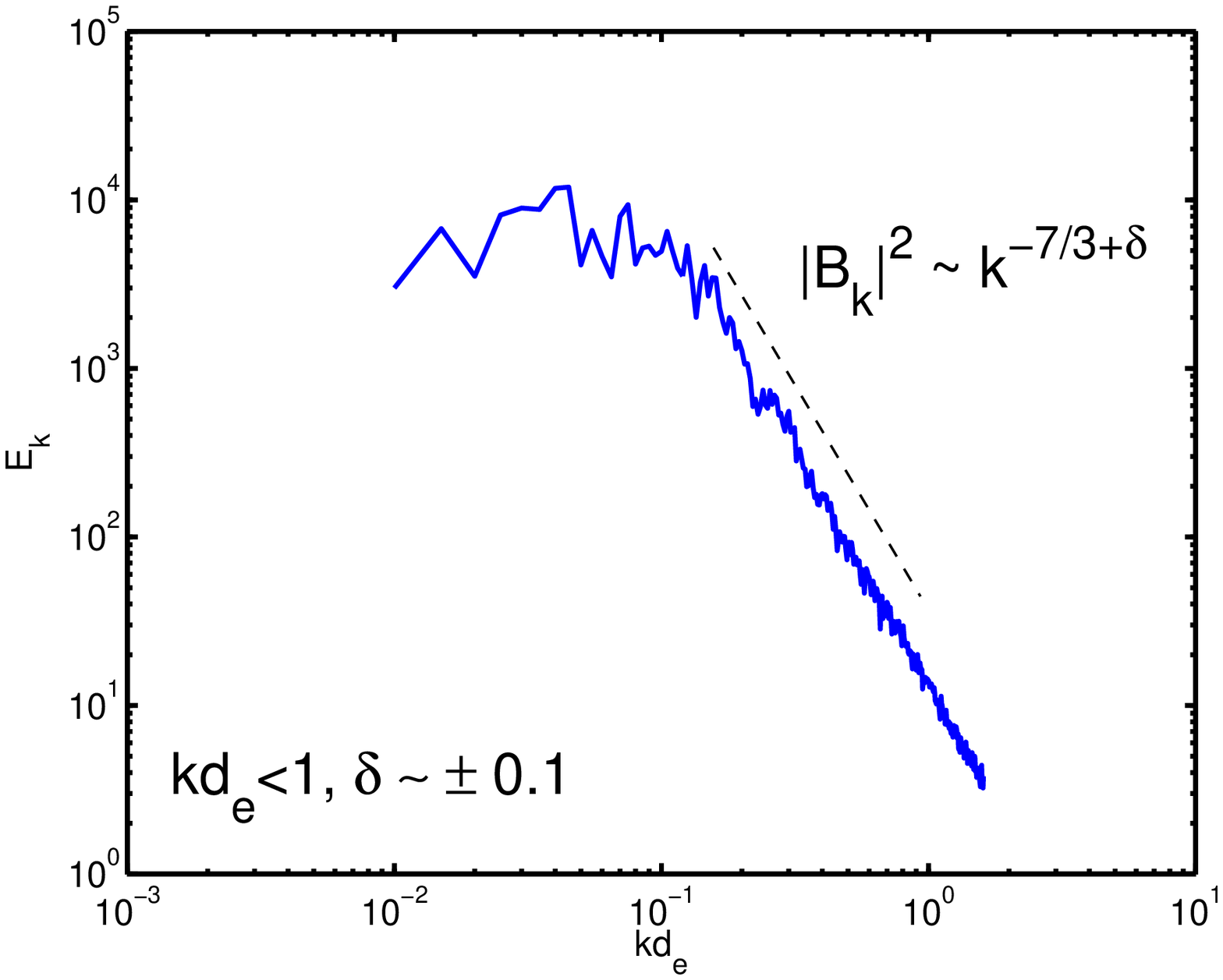,angle=0,width=0.5\textwidth}
\caption{\label{fig2a}} 
  Whistler  turbulence in the large scale $kd_e<1$ regime
  exhibits a Kolmogorov-like inertial range power spectrum close to
  $k^{-7/3}$. 
\end{figure}

\section{Energy spectra in whistler turbulence}
The spectral transfer of turbulent energy in the inertial range is
determined by neighboring Fourier modes in whistler turbulence.  We
find from our simulations that the mode coupling interaction follows a
Kolmogorov phenomenology \cite[]{kol, iros, krai} that leads to
Kolmogorov-like energy spectra. It is evident from \Fig{fig2a} that
whistler turbulence in the $kd_e<1$ regime exhibits a Kolmogorov-like
$k^{-7/3}$ spectrum. This inertial range turbulent spectrum, in the
context of low beta whistlers, is further consistent with previous 2D
work \cite[]{biskamp,dastgeer00a,dastgeer00b}.  Surprisingly, the
density fluctuations do not modify the energy spectrum in the $kd_e<1$
regime.  It turns out from the whistler wave dispersion relation that
the wave effects dominate in the large scale, i.e. $kd_e<1$, regime
where the inertial range turbulent spectrum depictes a Kolmogorov-like
$k^{-7/3}$ spectrum in our simulations.  Our previous work, on the
other hand, showed that the turbulent fluctuations in the smaller
scale ($kd_e>1$) regime behave like non magnetic eddies of
hydrodynamic fluid and yield a $k^{-5/3}$ spectrum (Shaikh 2009).  The
wave effect is weak, or negligibly small, in the latter. The observed
whistler turbulence spectra in the $kd_e<1$ regime in \Fig{fig2a} can
be followed from the Kolmogorov-like arguments \cite[]{kol, iros,
  krai} that describe the inertial range spectral cascades. We
elaborate on these arguments to explain our simulation results of
\Fig{fig2a} as follows.

\begin{figure}
\psfig{file=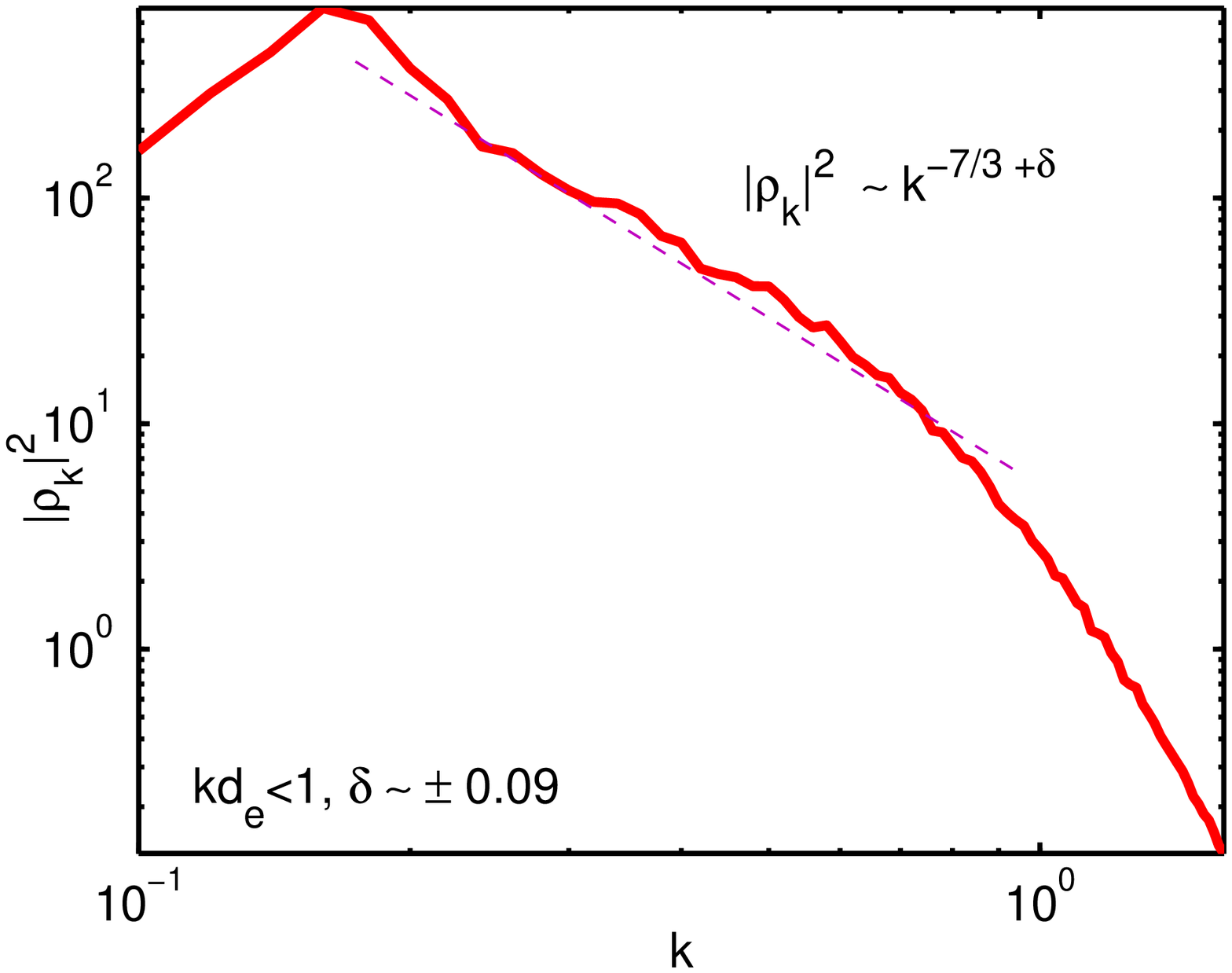,angle=0,width=0.45\textwidth}
\caption{\label{fig2b}} Spectrum of density fluctuations in whistler
 turbulence. The density fluctuations are slaved to the magnetic
field fluctuations and follow a Kolmogorov-like $k^{-7/3}$ spectrum in
the the $kd_e<1$ regime.
\end{figure}

In the low plasma-$\beta$ regime, the whistler turbulence model
described by the set of \eqsto{2demhd-eq1}{2demhd-eq4} admits the
following energy conservation law \cite[]{Abdalla}.  
\eqa 
\label{energy}
E = \frac{1}{2} \int dxdy
~[\phi^2 + (\nabla \psi)^2 + d_e^2 (\nabla \phi)^2 +  d_e^2 (\nabla^2\psi)^2 \nonumber \\
+ d_e \frac{\omega_{ce}}{\omega_{pe}}\{\beta^2 + \nabla \phi)^2 \}].
\eeq 
It is noted that the contribution due to the pressure and density
fluctuations does not modify the energy conservation relation in the
limit $\omega_{ce}/\omega_{pe} <1, ~{\bf B}/{B}_0<1$ and $\beta <1$.

The $kd_e<1$ regime comprises the dispersive whistler waves and the
total energy is dominated by first two terms in \eq{energy} such that
$E \simeq \phi^2 + (\nabla \psi)^2 $. Turbulent equipartition between
the velocity and magnetic field fluctuations yields $\phi \simeq
k\psi$ \cite[]{biskamp,dastgeer00a,dastgeer00b}. Owing thus to the
turbulent equipartition, the total energy can be given as $E \sim
\phi^2$ in the $kd_e<1$ regime.  The group velocity of whistler waves
in the $kd_e<1$ regime is $v_g \sim \partial \omega/\partial k_y \sim
k \sim \ell^{-1}$. Assuming that the nonlinear transfer of energy in
the inertial range is governed by the eddy interactions whose velocity
is $V_e \simeq \hat{z} \times \nabla \phi$. Applying Kolmogorov-like
dimensional arguments \cite[]{kol}, we obtain the $k^{\rm th}$ Fourier
component of the electron fluid velocity as $v_k \sim k \phi_k$. The
convective time scales on which the eddies transfer energy in the
inertial range can be estimated as $\tau_{nl} \sim 1/(kv_k) \sim
1/(k^2 \phi_k)$. The nonlinear energy cascade rates ($\varepsilon$)
are computed as $\varepsilon \sim E_k/\tau_{nl} \sim k^2 \phi_k^3$.

On using the Kolmogorov phenomenology that the spectral transfer is
local and depends only on the energy dissipation rates and modes
\cite[]{kol,iros}, the energy spectrum can be given by $E_k \sim
\varepsilon^\alpha k^\beta$. Upon substituting the energy dissipation
rates, we estimate the spectral energy as $E_k \sim \varepsilon^{2/3}
k^{-7/3}$. 

Thus, the energy spectrum $k^{-7/3}$ derived on the basis of
Kolmogorv-like arguments \cite[]{kol} is further consistent with our
simulations in \Fig{fig2a}.

\section{Effect of density fluctuations}
We find from our simulations that the contribution due to the pressure
and density fluctuations does not modify the energy conservation
relation in whistler turbulence. Hence the density fluctuations do not
influence the whistler wave cascades in the inertial range. This
result appears to be a bit surprising at first, but our understanding
is that the density fluctuations are only weakly coupled with the wave
magnetic field and hence they are slaved to the magnetic field. During
the evolution, density fluctuations simply follow the magnetic field
and exhibit a Kolmogorov-like energy spectrum.  The density spectrum
from our simulations is shown in \Fig{fig2b} for the $kd_e<1$ regime.

To quantitatively demonstrate that the density fluctuations couple
only weakly with the whistler wave magnetic field, we develop a novel
diagnostic to determine the contribution of density perturbation in
the whistler waves. This originates essentially from the whistler wave
relationship that relates the poloidal $\psi$ and axial $\phi$
components of the magnetic field flux functions and is described by
the expression $k|\psi_k|=|\phi_k|$
\cite[]{dastgeer00a,dastgeer00b,dastgeer05}. It should be further
noted that this relation is valid for the incompressible whistler mode
fluctuations. The effect of compressibility due to the density
perturbations however enters through \eqs{2demhd-eq3}{2demhd-eq4}.
The whistler relationship is thus modified to include the density
fluctuations and it reads as
\be
k^2|\psi_k|^2= \left|\frac{n_k}{\lambda^2 k^2} -\beta_k \right|^2,
\ee
where the left hand side corresponds to the energy associated with
the wave field, whereas the right hand side of the expression
describes the energy associated with the density field.
Based on the modified whistler relationship, we develop a parameter
called as {\em whistler parameter} in the following.
\be
\label{prm}
\chi(t) = \frac{\sum_{k_x}\sum_{k_y} \left(k^2|\psi_k|^2- \left|\frac{n_k}{\lambda^2 k^2} -\beta_k \right|^2 \right)}
{\sum_{k_x}\sum_{k_y} \left( k^2|\psi_k|^2+ \left|\frac{n_k}{\lambda^2 k^2} -\beta_k \right|^2 \right)}.
\ee

The whistler parameter is a dimensionless quantity that determines the
contribution of density fluctuations in the whistler waves over the
entire 2D turbulent spectrum. The physical picture emerging from this
parameter can be described as follows; During the evolution of
whistler waves and the corresponding density field, if there is a
signicant amount of energy being transferred in the density
fluctuations (i.e. comparable to the whistler wave energy), then the
difference in the magnitude of the wave energy $k^2|\psi_k|^2$ and
density fluctuations $|n_k/\lambda^2 k^2 -\beta_k|^2$ will be
minimal. Therefore a normalized contribution, obtained by dividing the
difference by the total energy in the wave and the density field i.e.
$\sum_{k_x}\sum_{k_y} (k^2|\psi_k|^2+ |n_k/\lambda^2 k^2
-\beta_k|^2)$, will be much smaller than unity. This essentially
corresponds to a state where $\chi \ll 1$. Such a criterion further
characterizes a state in which the whistler wave magnetic field is
greatly influenced or contaminated by the density fluctuations. By
contrast, the limit $\chi \rightarrow 1$ corresponds to a state in
which density fluctuations contribute {\em only weakly} or negligibly
small in the wave magnetic field.

\begin{figure}
\psfig{file=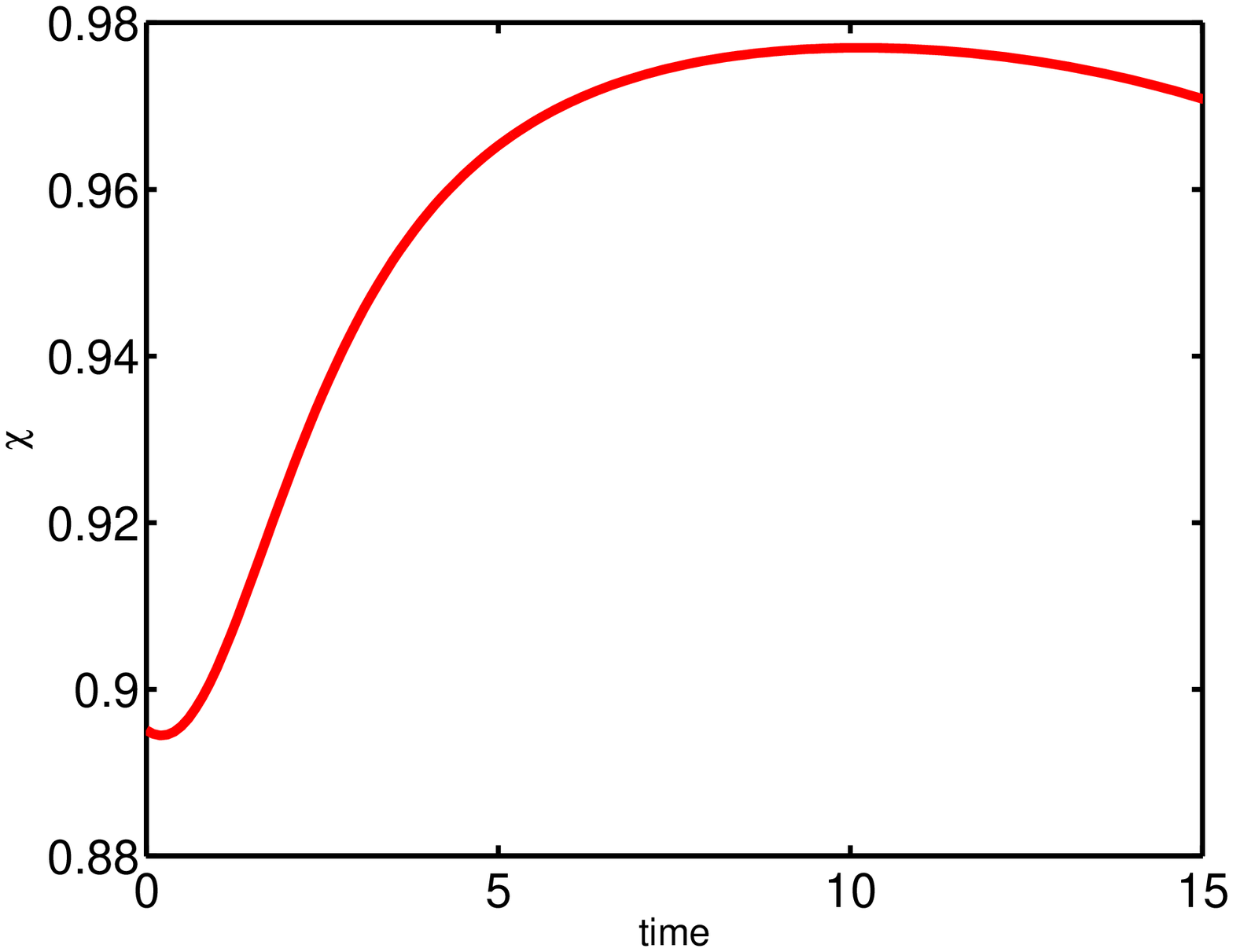,angle=0,width=0.45\textwidth}
\caption{\label{chi}} Evolution of whistler parameter $\chi$ that
determines the contribution of density fluctuations in the whistler
waves in the $kd_e<1$ regime. The parameter $\chi$ approaching unity
corresponds to the whistler wave dominated state in which density
fluctuations are weakly coupled and they do not influence the energy
cascade processes.
\end{figure}

We follow the evolution of $\chi$ in our simulations. The result is
plotted in \Fig{chi}. It is evident from \Fig{chi} that $\chi$ is
increasing progressively and it is approaching unity i.e.  $ \chi
\rightarrow 1$. This means that energy associated with the wave field
dominates over the energy in the density fluctuations. A decreasing
trend of $\chi$, on the other hand, would have corresponded to state
in which the energy in the density field was dominated over the
wave. Since our simulations indiate that $\chi \rightarrow 1$, we
believe that the effect of density perturbations is rather weak on the
evolution of whistler waves and hence energy cascade rates are not
altered. It is because of this reason that we find a Kolmogorov-like
$k^{-7/3}$ energy spectrum in our simulations, a result similar to the
one obtained previously by \cite{biskamp,biskamp2,dastgeer05,cho1}.

Although Kolmogorov-like $k^{-7/3}$ energy spectrum has been reported
previously \cite[]{biskamp,biskamp2,dastgeer05,cho1}, they ignored the
effect of compressibility due to the density field in whistler
turbulence.  Our work described here is therefore different in a sense
that we have included the density fluctuations in whistler turbulence
and found that they do not influence the energy cascade processes in
the inertial range spectrum. The inertial range turbulent spectrum in
the $kd_e<1$ regime thus exhibits a Kolmogorov-like $k^{-7/3}$ power
law. It thus follows from \fig{fig2b}{chi} that the density field is
simply slaved to the magnetic field and leads to Kolmogorov-like
spectrum in the wave-dominated $kd_e<1$ regime. Moreover, it should be
noted from \Fig{chi} that the evolution of whistler parameter is
dominated entirely by the wave energy (over the density field) right
from the outset. A primary reason of such dominance is ascribed to the
form of \eq{2demhd-eq4} that relates density to the small scale
density and potential fluctuations. This essentially means that the
density field is dominated by the characteristic small scale
fluctuations.  Such small scale fluctuations tend to possess a weaker
tendency to influence the turbulent spectrum in the $kd_e<1$ regime
where the large scale whistler waves govern the entire nonlinear
physics. This description further leads to another plausible
explanation with regard to the time scales.  The large scale energy
containing whistler modes evolve on slower time scales
[c.f. \eq{disp}]. By contrast, the small scale density fluctuations
evolve on faster time scales. Owing to this temporal disparity
associated with the two, the density field does not spend enough time
with the wave field to influence its dynamical evolution. Hence the
former is substantially incapable of coupling with the wave field.
This leads to the negligibly small contribution of the density
fluctuations in the wave field.  The inertial range cascade is
therefore governed predominantly by the whistler mode interactions.

\section{Anisotropic compressible whistler  turbulence}
\begin{figure}
\psfig{file=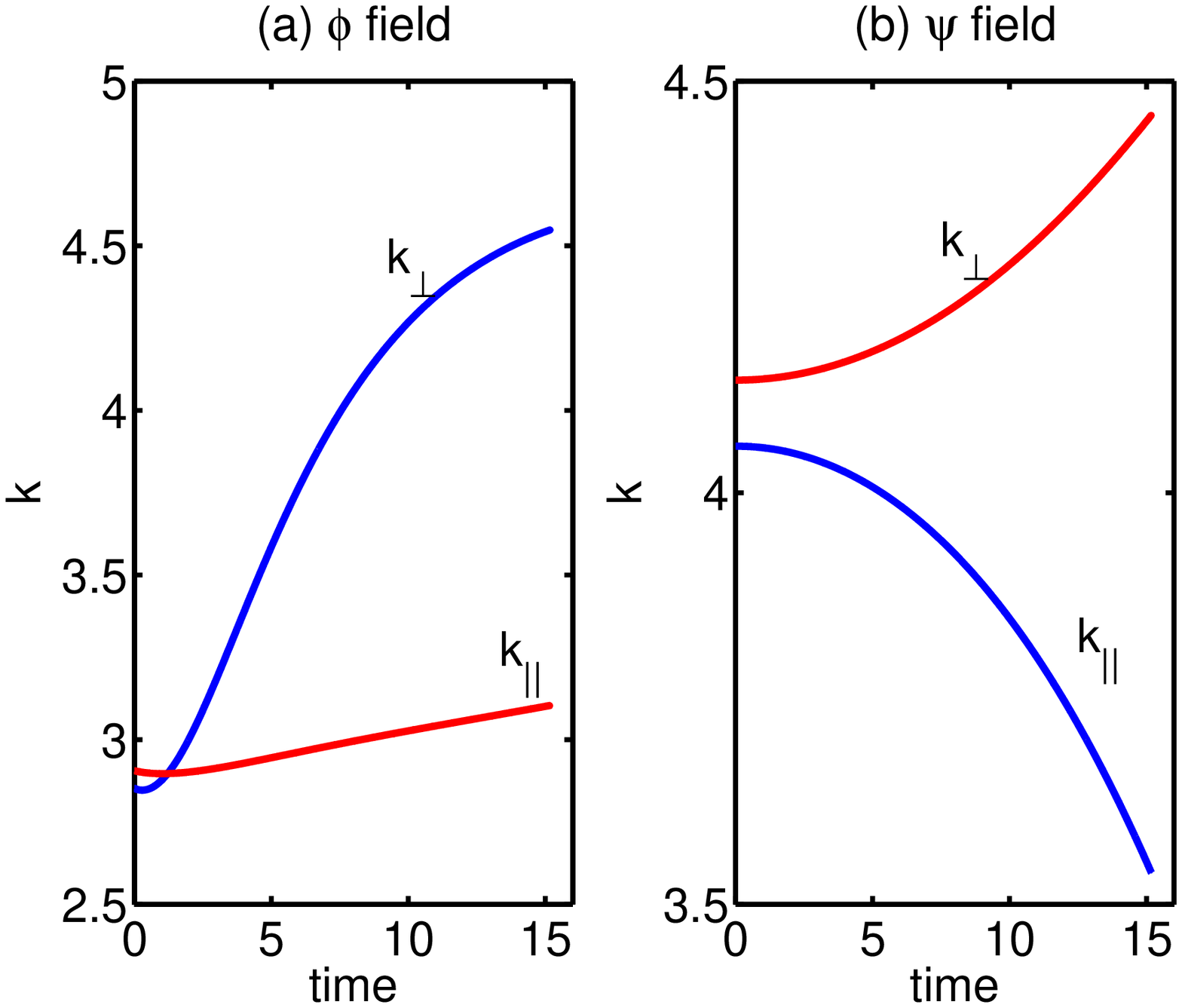,angle=0,width=0.5\textwidth}
\caption{\label{fig3}} 
Evolution of anisotropic modes $k_\parallel$ and $k_\perp$ averaged
over the entire turbulent spectrum in low beta whistler
turbulence. Initially, $k_\parallel \simeq k_\perp$. Progressive
development of anisotropy $k_\parallel < k_\perp$ is ascribed to the
presence of background magnetic field.
\end{figure}

We next quantify the degree of anisotropy mediated by the presence of
large scale magnetic in the nonlinear 2D whistler turbulence. In 2D
turbulence, the anisotropy in the $k_x-k_y$ plane is associated with
the preferential transfer of spectral energy that empowers either of
the $k_x$ and $k_y$ modes. The background magnetic field is considered
along the $x$ direction in our simulations. We therefore expect
asymmetry in the evolution of the $k_x$ and $k_y$ modes.  The
anisotropy in the initial isotropic turbulent spectrum is triggered
essentially by the background large scale magnetic field that regulate
turbulent fluctuations to nonlinearly migrate the spectral energy in a
particular direction. To measure the degree of anisotropic cascades,
we employ the following diagnostics to monitor the evolution of
$k_\parallel$ mode in time. The $k_\parallel$ mode is determined by
averaging over the entire turbulent spectrum that is weighted by $k_x$
which is aligned in the direction of the background magnetic field.
\be
\label{kpara}
k_\parallel(t) = \sqrt{\frac{\sum_k |k_x Q(k,t)|^2}{\sum_k |Q(k,t)|^2}}
\ee
Here $Q$ represents any of $\phi$, $\psi$, $\nabla^2 \psi$ and
$\nabla^2 \phi$.  Similarly, the evolution of $k_\perp$ mode across
the is background magnetic field determined by the following relation.
\be
\label{kperp}
k_\perp(t) = \sqrt{\frac{\sum_k |k_y Q(k,t)|^2}{\sum_k |Q(k,t)|^2}}
\ee
It is clear from these expressions that the $k_x$ and $k_y$ modes
exhibit isotropy when $k_x \simeq k_y$.  Any deviation from this
equality leads to a spectral anisotropy. We follow the evolution of
$k_\parallel$ and $k_\perp$ modes in our simulations for long enough
time. Our simulation results describing the evolution of $k_\parallel$
and $k_\perp$ modes are shown in \Fig{fig3}. It is evident from
\Fig{fig3} that the initial isotropic modes $k_x \simeq k_y$ gradually
evolve towards an highly anisotropic state in that spectral transfer
preferentially occurs in the $k_\perp$ mode, while the same is
suppressed in $k_\parallel$ mode. Consequently, the spectral transfer
in $k_\perp$ mode dominates the evolution and the mode structures show
elongated structures along the $x$-direction, see \Fig{fig1}.

The evolution of $k_\perp$ and $k_\parallel$ shows a significant
disparity in the two modes by virtue of an external magnetic field.
The evolution of $k_\perp$ and $k_\parallel$ is clearly different as
the spectral cascade in the parallel wavenumbers is dramatically
suppressed.  The suppression is caused by the excitation of whistler
waves, which act to weaken spectral transfer along the direction of
propagation. This can be understood as follows; We assume that the
spectral transfer, essentially mediated by propagating whistler waves
in wavenumber space, can be described by a three wave interaction
mechanism, for which the frequency and wavenumber resonance criteria
are, respectively, expressed by
\[ \pm \omega_3 = \omega_1 - \omega_2, \]
and
\[ {\bf k_3} = {\bf k_1} +{\bf k_2}. \]
The resonance conditions indicate that two whistler waves
$(\omega_1,{\bf k_1})$ and $(\omega_2,{\bf k_2})$ mutually interact
and give rise to the third wave $(\omega_3,{\bf k_3})$. Such
conditions could, in principle, hold for a set of infinite waves as
the indices `$1$' and `$2$' are merely dummy indices. With the help of
dispersion relation (say, in the $kd_e<1$ regime) and using the
wavenumber $k_{3_y}=k_{1_y}+ k_{2_y}$, we can obtain
\[ \frac{k_{1_y}}{k_{2_y}} = \frac{k_{2}+k_{3}}{k_{1}-k_{3}}. \]
Let us now suppose the Kolmogorov turbulence hypothesis holds for EMHD
nonlinear interactions as well, viz, that spectral transfer in the
wavenumber space is local and occurs efficiently amongst the most
adjacent Fourier modes, i.e.  for which $|k-k_1| \approx |k_1|$. It,
then, implies 
\be k_{2_y} < k_{1_y}
\label{wt}
\ee 
since $k_{1} \approx k_{2} \approx k_{3}$, thereby indicating that
there is a very little cascading along the $\hat{y}$-direction
i.e. the magnetic field direction. Thus, the parallel wavenumbers
($k_\parallel$) appear to be suppressed and the spectral cascade
mainly occurs in the perpendicular wavenumbers ($k_\perp$). This, we
suggest, explains the wavenumber disparity ($\langle k_\perp \rangle
\ne \langle k_\parallel \rangle$) observed in our simulations [see
  \Fig{fig3}].

\section{Discussion and Conclusions}
In summary, we present the results of freely decaying whistler
turbulence calculated from an EMHD computer model.  A major emphasis
is to understand the effect of density fluctuations on the inertial
range cascades in the whistler wave dominated $kd_e<1$ regime. In this
regime, large scale whistler waves govern the energy cascade
processes.  Interestingly we find that despite strong density
perturbations, their effect on the cascade dynamics is
inconsequential. Hence they couple weakly with the large scale waves
in the $kd_e<1$ regime. We find from our simulations that the density
fluctuations do not influence the inertial range turbulent
spectra. Consequently, the turbulent fluctuations in the inertial
range are described by Kolmogorov-like phenomenology. Thus consistent
with the Kolmogorov-like dimensional argument, we find that turbulent
spectra in the $kd_e<1$ regime is described by $k^{-7/3}$. Our results
are important particularly in understanding turbulent cascade
corresponding to the high frequency ($\omega > \omega_{ci}$) solar
wind, space and astrophysical plasmas where characteristic
fluctuations are comparable to the electron inertial skin depth. Our
work might also be useful in describing why many space plasmas are
described by by the Kolmogorov-like energy spectra despite the
presence of strong density fluctuations and/or compressible effects.
We find that density fluctuations exist on smaller scales and have
typically higher frequency associated with their evolution. Owing to
these disparate length and time scales, they do not modify the large
scale wave-dominated turbulent processes. It is this large scale
dynamics in the $kd_e<1$ regime that leads to the Kolmogorov-like
$k^{-7/3}$ spectrum in our simulations.

Note that our simulations of inhomogeneous whistler turbulence
pertains to decaying turbulence only.  In principle, turbulence can be
driven. It remains to be seen whether the energy cascade processes in
the driven whistler turbulence are altered by the density
fluctuations.

\section*{Acknowledgments}
The support of NASA(NNG-05GH38) and NSF (ATM-0317509) grants is
acknowledged.


\vfill\eject

\end{document}